\documentclass[reprint,prb,superscriptaddress,showpacs]{revtex4-2}%
\usepackage{graphicx}
\usepackage{color}
\usepackage{amsmath}
\usepackage{amsfonts}
\usepackage{amssymb}
\usepackage{framed}
\usepackage{float}
\providecommand{\U}[1]{\protect\rule{.1in}{.1in}}
\makeatletter
\renewcommand*{\fnum@figure}{{\normalfont\bfseries \figurename~\thefigure}}
\renewcommand*{\@caption@fignum@sep}{\textbf{ : }}
\makeatother
\usepackage{hyperref}
\hypersetup{
    colorlinks=true,
    linkcolor=blue,
    filecolor=magenta,      
    urlcolor=cyan,
}
\begin{document}
\title{ CoTe$_2$: A quantum critical Dirac metal with strong spin fluctuations}
\author{Peter E. Siegfried}
\affiliation{Department of Physics and Astronomy, George Mason University, Fairfax, VA 22030, USA.}
\affiliation{Quantum Science and Engineering Center, George Mason University, Fairfax, VA 22030, USA.}
\author{Hari Bhandari}
\affiliation{Department of Physics and Astronomy, George Mason University, Fairfax, VA 22030, USA.}
\affiliation{Quantum Science and Engineering Center, George Mason University, Fairfax, VA 22030, USA.}
\author{Jeanie Qi}
\affiliation{Thomas Jefferson High School, Alexandria VA 22312, USA}
\author{Rojila Ghimire}
\affiliation{Central Department of Physics, Tribhuvan University, Kirtipur, Kathmandu 44613, Nepal.}
\author{Jayadeep Joshi}
\affiliation{Department of Physics and Astronomy, George Mason University, Fairfax, VA 22030, USA.}
\address{Quantum Science and Engineering Center, George Mason University, Fairfax, VA 22030, USA.}
\author{Zachary T. Messegee}
\affiliation{Department of Chemistry and Biochemistry, George Mason University, Fairfax, Virginia 22030, USA.}
\author{Willie Beeson}
\affiliation{Physics Department, Georgetown University, Washington, DC 20057, USA.}
\author{Kai Liu}
\affiliation{Physics Department, Georgetown University, Washington, DC 20057, USA.}
\author{Madhav Prasad Ghimire}
\affiliation{Central Department of Physics, Tribhuvan University, Kirtipur, Kathmandu 44613, Nepal.}
\author{Yanliu Dang}
\affiliation{Materials Science and Engineering Division, National Institute of Standards and Technology (NIST), Gaithersburg, Maryland 20899, USA.}
\author{Huairuo Zhang}
\affiliation{Materials Science and Engineering Division, National Institute of Standards and Technology (NIST), Gaithersburg, Maryland 20899, USA.}
\affiliation{Theiss Research, Inc., La Jolla, California 92037, USA.}
\author{Albert Davydov}
\affiliation{Materials Science and Engineering Division, National Institute of Standards and Technology (NIST), Gaithersburg, Maryland 20899, USA.}
\author{Xiaoyan Tan}
\affiliation{Department of Chemistry and Biochemistry, George Mason University, Fairfax, Virginia 22030, USA.}
\author{Patrick M. Vora}
\affiliation{Department of Physics and Astronomy, George Mason University, Fairfax, VA 22030, USA.}
\affiliation{Quantum Science and Engineering Center, George Mason University, Fairfax, VA 22030, USA.}
\author{Igor I. Mazin} 
\affiliation{Department of Physics and Astronomy, George Mason University, Fairfax, VA 22030, USA.}
\affiliation{Quantum Science and Engineering Center, George Mason University, Fairfax, VA 22030, USA.}
\author{Nirmal J. Ghimire}
\email{Corresponding author: nghimire@gmu.edu}
\affiliation{Department of Physics and Astronomy, George Mason University, Fairfax, VA 22030, USA.}
\affiliation{Quantum Science and Engineering Center, George Mason University, Fairfax, VA 22030, USA.}

\date{\today}
%*********************************************************************************************************************************
\begin{abstract}
Quantum critical points separating weak ferromagnetic and paramagnetic phases trigger many novel phenomena. Dynamical spin fluctuations not only suppress the long-range order, but can also lead to unusual transport and even superconductivity. Combining  quantum criticality with topological electronic properties presents a rare and unique opportunity. Here, by means of ab initio calculations and magnetic, thermal, and transport measurements, we show that the orthorhombic CoTe$_2$ is close to ferromagnetism, which appears suppressed by spin fluctuations. Calculations and transport measurements reveal nodal Dirac lines, making it a rare combination of proximity to quantum criticality and Dirac topology.
 \end{abstract}
 \maketitle
 
\section*{Introduction}
In condensed matter physics, interplay between ionic, electronic, and magnetic degrees of freedom can generate many novel phenomena and quasiparticle excitations.
Of particular interest have been topologically  protected (i.e., by symmetry) band crossings and other electronic features. The field started with the study of topological insulators, then expanded to semimetals, first to 3D Dirac and Weyl points \cite{Armitage2018} and even to metals \cite{Burkov2011, Kim2015, Bzdusek2016, Rhim2015b,Zhu2016,Ahn2018,Ghimire2018f,Khoury2019}, leading to unusual magnetotransport properties. The novel physics and rarity of these states, as well as their potential applications \cite{Gilbert2021,Polash2021,Kumar2021} have been at the forefront of recent research.
This said, more often than not it is difficult to see the effect of topologically nontrivial points in metals because they tend to be shadowed by other, trivial parts of the Fermi surface.
In this connection, a search has been underway for metals that possess symmetry-driven planes of degeneracy in the momentum space, which can cross the Fermi level and generate continuous nodal lines of Dirac points directly at the Fermi level \cite{Leonhardt2021}. Such metals would have a dramatically larger part of the Fermi surface bearing topological phenomena, and thus have potential for more unusual properties such as 3D quantum anomalous Hall effect \cite{Rhim2015b,Xu2011}, tunable Weyl points \cite{Yan2016}, and ``drumhead'' surface states \cite{Bian2016} that are argued to provide a route to higher temperature superconductivity \cite{Kopinn2011}.

Coexistence or proximity of such topological states with other exotic phenomena such as non-trivial spin textures,  superconductivity, or quantum criticality can further provide a new route to the properties expected from the interplay of real and momentum space topology, originally discussed in the A-phase of $^{3}$He \cite{Volovik2003a,Volovik2007}. These promising theoretical proposals have driven and guided recent experimental efforts toward the realization and study of such materials.

Quantum criticality, in particular, zero-temperature magnetic phase transitions, is believed to have connections to several emergent phenomena such as  triplet superconductivity \cite{Singh2002}, non-Fermi liquid scalings \cite{Stewart2002}  and unusual  magnetotransport behavior \cite{Mazin2004}. Some famous examples of such materials are unconventional superconductors (Sr$_2$RuO$_4$ and Fe-based superconductors), skyrmionic materials (MnSi and FeSi) and weak itinerant ferromagnet, and superparamagnets (ZrZn$_2$, Ni$_3$Al, Ni$_3$Ga, Pd). As such, a quantum critical system with a substantial presence of topologically-nontrivial states at the Fermi level would provide a novel materials platform for the study of the interplay of real and momentum space topology, but such materials are quite rare. 

In this work, we report the synthesis, and magnetic, thermal and magnetotransport properties of single crystal CoTe$_2$,  and argue that it is a quantum critical Dirac metal with   Dirac points and multiple nodal lines at the Fermi surface. The material does not show magnetic ordering down to 1.8 K despite the DFT ground state being distinctly ferromagnetic, a hallmark of a magnetic order suppressed by quantum-critical spin fluctuation. We find that CoTe$_2$ is well described by 
Moriya's Self-Consistent Renormalization Theory (SCRT) \cite{Moriya1985}, further supporting the conclusion that CoTe$_2$ is close to a ferromagnetic quantum critical point. This is highly interesting for its potential for tuning magnetic correlations in the presence of a unique topological electronic behavior. 

 \section*{Results and Discussion}
Single crystals of CoTe$_2$ were grown using the Te self flux method as described in the Methods section. CoTe$_2$ has previously been reported in three different space groups \cite{Brostigen1970,Cheda1986,Ma2018} so the structure of our bulk crystals was verified using multiple complementary methods: powder and single crystal x-ray diffraction, atomic resolved scanning transmission electron microscopy (STEM), and Raman spectroscopy, all confirming the orthorhombic marcasite structure in  space group Pnnm(\#58) [see  Supplementary Materials (SM) sections S1 to S3 for a full structural description] \cite{NIST}. 

\begin{figure}[ht!]
\begin{center}
\includegraphics[width=1\linewidth]{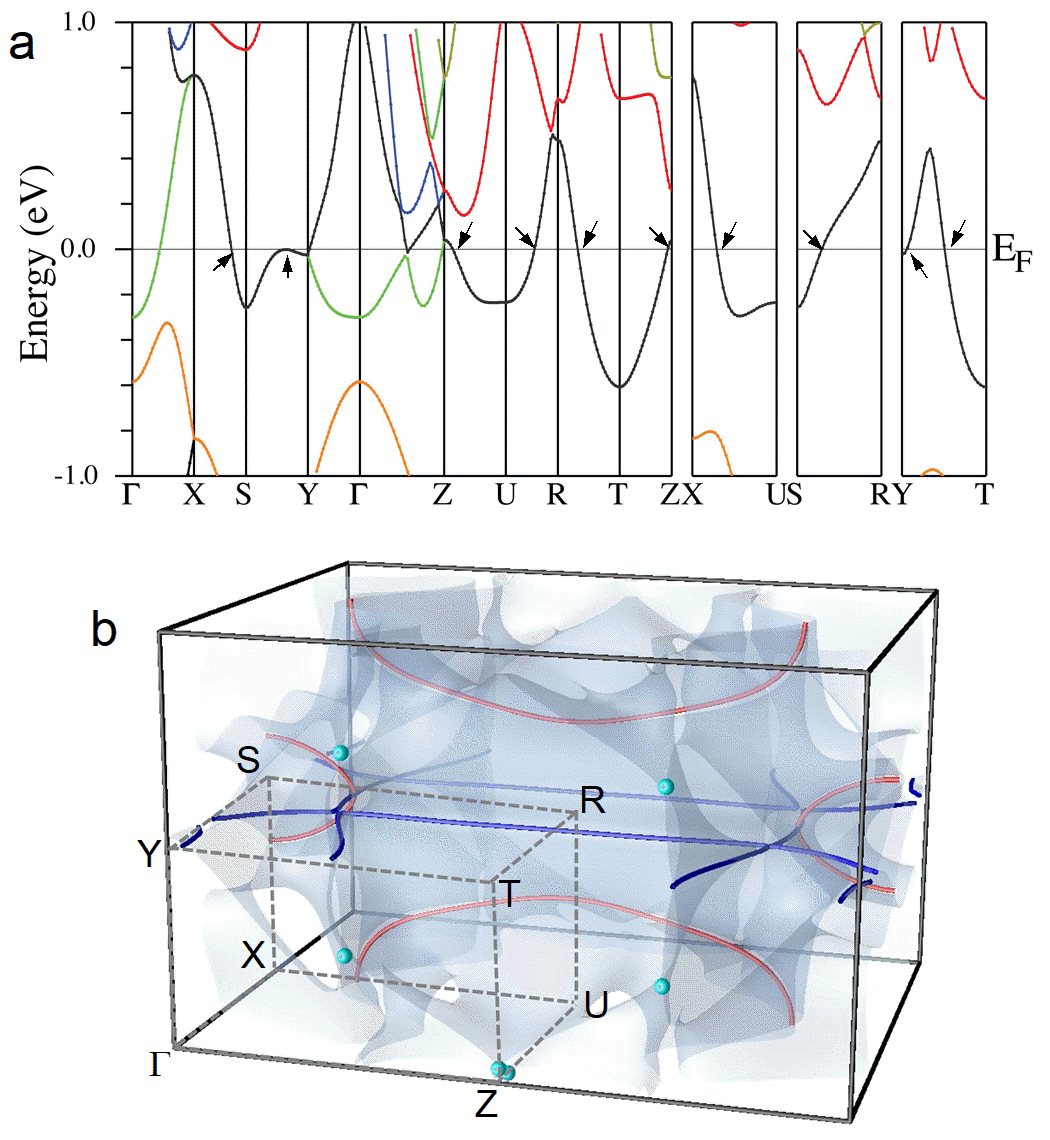}
\end{center}
\caption{(a) Calculated band structure for CoTe$_2$. The arrows indicate locations of the Fermi-level Dirac points on these lines, and directions of linear dispersion (see more in SM section S4). (b) The Fermi surface in the first reciprocal lattice cell (from $\Gamma$ to $\Gamma$ in all directions) with red and blue tubes showing two types of the the Dirac nodal lines and cyan balls indicating Dirac points (except near the top Z point).}
\label{F1}
\end{figure}

We first start by summarizing the electronic band structure of CoTe$_{2}.$ Fig.~\ref{F1} shows the scalar-relativistic (no spin-orbit coupling) band
structure along high-symmetry directions and the Fermi surface. 
The former, depicted in Fig. \ref{F1}(a), shows a surprisingly large number of degenerate bands, namely along
the X-S-Y, Z-U-R-T-Z, X-U, S-R and Y-T lines. To understand this, we recall
that CoTe$_{2}$ crystallizes in an interesting orthorhombic symmetry, Pnnm
(\#58). This symmetry group is highly asymmorphic: of its 8 symmetry
operations half are glide planes. As a result, it has a large number of nodal
lines and planes, where all electronic states are doubly degenerate by
symmetry. Leonhard $et$ $al.$ analyzed all orthorhombic groups \cite{Leonhardt2021} and
concluded that for this group there are two such planes, defined by the
conditions $k_{x}=\pi/a$ and $k_{y}=\pi/b,$ respectively, and two lines, Z-T
and Z-U. Degenerate planes generate nodal lines on the Fermi surface, which in
CoTe$_{2}$ form two closed loops [shown in Fig.  \ref{F1}(b) as red
loops] in the $k_{x}=\pi/a$ plane and two sets of intersecting infinite lines,
plus a small loop around Y (blue) in the $k_{y}=\pi/b$ one. The two lines
identified in Ref. \cite{Leonhardt2021} generate two nodal points, shown as two cyan
balls near Z in the same figure (note that while these form a quartet of four
degenerate points near Z, they do not form, in agreement to the theoretical predictions \cite{Leonhardt2021}, a closed loop). To the best of our knowledge this is the first
direct verification of the Ref. \cite{Leonhardt2021}'s predictions for the $Pnnm$ space group.
Note that, in agreement with Ref. \cite{Leonhardt2021}, the degeneracy is lifted
linearly in all directions, providing a large phase space for the Dirac electrons.

In addition, there is a line of \textit{accidental} degeneracy, where two
particular bands of Co origin are allowed to cross (without hybridization) in
the $k_{x}=0$ plane. This line intersects with the Fermi surface at a Dirac
point that sits close (but not exactly at) the center of the $\Gamma$-Y-T-Z
rectangle, and is also shown in Fig. \ref{F1}(b) by a cyan ball. This Dirac point
is 3D (the degeneracy is lifted linearly in all directions) and strongly tilted. 

Last but not least, there is one other special Dirac point in the system. From the topological (in classical sense) point of view an isolated Dirac crossing, by definition, only shows degeneracy at one single point in the Brillouin zone, which can only appear on the Fermi level by accident (DP-0). The next level is a nodal line, i.e., a line of Dirac points, which can cross the Fermi surface at isolated points (DP-1). An even stronger, and much less common, there is a nodal plane generating a line of Dirac points on the Fermi level (DP-2). In our case, we have all of those, plus a special Dirac point on the Fermi surface between S and R. Here, the Dirac degeneracy is protected not only when one component of the wave vector is changing (as in DP-1), or two components (DP-2), but all three components $k_x$, $k_y$, and $k_z$ (DP-3). Only when two or more components are varied simultaneously, the degeneracy is lifted (see SM section S4 and Figs. S4 and S5 for details). This is the ``strongest" Dirac point that can exist in a 3D system, and it occurs rather rarely in real systems.
 
\begin{figure}[!ht]
\begin{center}
\includegraphics[width=1\linewidth]{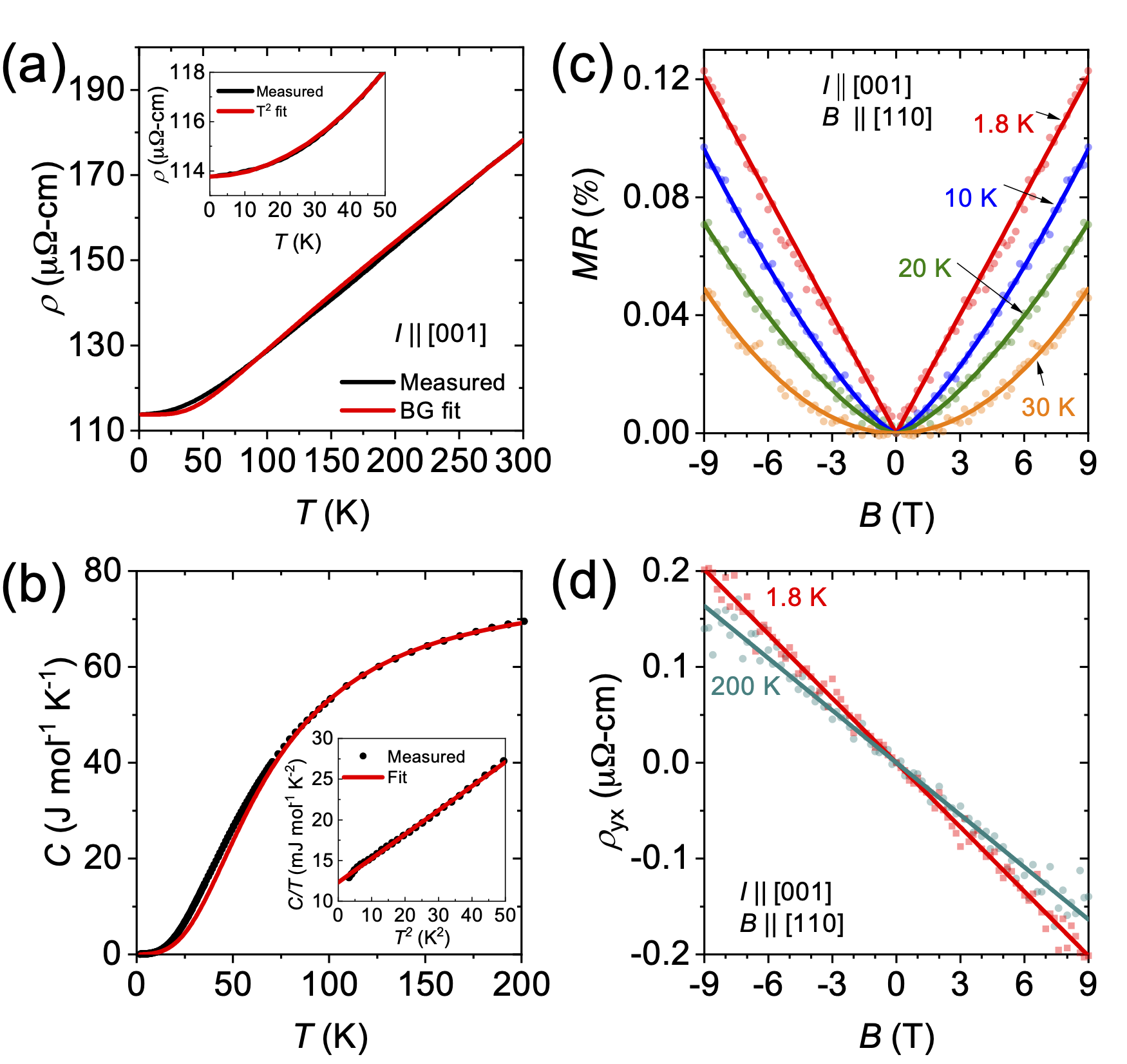}
\end{center}
\caption{\protect\small(a) Temperature dependence of electrical resistivity (black) and Bloch-Gr{\"u}neisen fit (red) of CoTe$_2$ measured with current applied along [001]. Inset shows low-temperature data fitted to $\rho_0$+A$T^2$; $\rho_0$=113.8 $\mu\Omega$-cm, and A = 0.00173. (b) Specific heat capacity of CoTe$_2$ as a function of temperature (black) and Debye model fit (Eq. \ref{FullC}, red). The inset shows the measured $T<$7 K $C/T$ vs $T^2$ dependence (black) and linear fitting to Eq. \ref{lowTC} (red). (c) Magnetoresistance (MR) of CoTe$_2$ at indicated temperatures. Solid lines are fits of different functions to the data as described in the text. (d) Hall resistivity vs magnetic field at indicated temperatures. Solid lines are straight line fits to the data.}
\label{F2}
\end{figure}

Often, presence of the Dirac bands near the Fermi surface gives rise to unusual transport signatures. In clean semimetals, they may give rise to extremely large magnetoresistance (MR), quantum oscillations and non-zero Berry phase \cite{Ali2014,Ghimire2015,Liang2015,Ghimire2016a,Burkov2014,Murakawa2013} (it is to be noted that the extremely large MR and quantum oscillations  in semimetals can have also a non-topological origin \cite{Luo2016c,Sun2016a,He2016}). However, a combination of multiple Dirac points and lines on the Fermi surface (coming from nodal lines and planes passing through the entire BZ) have rarely been observed and their manifestation in the transport measurements have not been discussed. To this end, we carried out transport measurements on the single crystals of CoTe$_2$. Electrical resistivity of CoTe$_2$ as a function of temperature measured with current ($I$) along [001] is depicted in Fig. \ref{F2}(a) (black curve), which shows a metallic character in the entire temperature range between 1.8 and 300 K, consistent with the calculated electronic structure, and does not show any feature indicative of long range magnetic ordering.  Specific heat capacity from 1.8 to 200 K is shown in Fig. \ref{F2}(b) (black curve), which shows a monotonic change with temperature not showing any sign of long range ordering as in the resistivity measurement. Although no long range magnetic ordering is observed in resistivity and heat capacity measurements, their analysis provide information on the average phonon frequency in terms of the Debye temperature, $\mathrm{\Theta_D}$.  The Debye formula for the specific heat as a function of temperature is
\begin{equation}
C(T)=9R\left(\frac{T}{{\Theta_\mathrm D}}\right)^3\int_0^{{\Theta_ \mathrm D}/T}\frac{t^4e^t}{(e^t-1)^2}{d}t + \gamma T, 
\label{FullC}
\end{equation}
where the first term gives the phonon contribution to the specific heat, and $R$ is the gas constant. The second term is the Sommerfeld electronic specific heat with the coefficient $\gamma$ giving the mass-renormalized density of states. In the low temperature limit this simplifies to,
\begin{equation}
C(T)=AT^3 + \gamma T. 
\label{lowTC}
\end{equation}
where the first term is the contribution from acoustic phonons, $A={12\pi^4R}/{5{\Theta_ \mathrm D}^3}$. Fitting both the full temperature range to Eq. \ref{FullC} [red curve in Fig. \ref{F2}(b)], as well as just the low temperature data to Eq. \ref{lowTC} [inset in Fig. \ref{F2}(b)] yield similar results finding $\mathrm{\Theta_D}$= 275.4 K and $\gamma= 14.22$ mJ$\cdot$mol$^{-1}$K$^{-1}$ for Eq. \ref{FullC} and $\mathrm{\Theta_D}= 273.1$ K and $\gamma= 12.45$ mJ$\cdot$mol$^{-1}$K$^{-1}$ for Eq. \ref{lowTC} and $T< 7$ K, indicating that the contribution from optical phonons at lower temperatures is small. This  $\gamma$ is to be compared with the Sommerfeld coefficient from the unrenormalized DFT density of states, $N(E_F)\approx 6.5$ states/eV$\cdot$f.u., $\gamma_0\approx 7.6$ mJ$\cdot$mol$^{-1}$K$^{-1}$, giving a mass renormalization between 60 and 90\%; note that if this renormalization were due to electron-phonon coupling the material would have been a good superconductor, so we interpret this as the manifestation of critical spin fluctuations discussed below. 

We further analyzed the temperature dependence of the resistivity in terms
of the Bloch-Gr{\"u}neisen (BG) formula\cite{Ziman2001}, where the electrical resistivity is
\begin{equation}
\rho_{xx}(T)=\rho_0+\rho_{ph}(T)
\end{equation}
and $\rho_0$ is the residual resistivity at $T=0$. The temperature-dependent term is given by\cite{Ziman2001,Bid2006}, \begin{equation}
\rho_{ph}(T)=\alpha_n\rho_{{\Theta_\mathrm R}}\left(\frac{T}{{\Theta_\mathrm R}}\right)^n\int_0^{{\Theta_\mathrm R}/T}\frac{t^n}{(e^t-1)(1-e^{-t})}{d}t,
\label{n5}
\end{equation}
where $\alpha_n$ is a coefficient describing the scattering rate, $\mathrm{\Theta_R}$ is the BG temperature, which is typically close, albeit not always equal to the Debye temperature, $\rho_{\mathrm{\Theta_R}}$ is the residual resistivity at $T=\mathrm{\Theta_R}$, and the exponent $n$ takes the value of 2, 3, or 5 depending on the specifics of the electronic interactions present within the material \cite{Cvijovic2011,Wilson1938,Barber1937}. For standard acoustic phonon scattering, the exponent $n$ is 5 and the coefficient $\alpha_{5}$ is 4.225. Inserting these values for acoustic phonons as well as fixing the $\mathrm{\Theta_R}=\mathrm{\Theta_D}$ of 275.4 K, yields a temperature dependent resistivity in close agreement with experiment. It is known that the resistivity can observe a $T^2$ dependence in the presence of high paramagnetism and large electronic specific heat \cite{Barber1937}, and further, the assumption that $\mathrm{\Theta_R}=\mathrm{\Theta_D}$ isn't strictly true \cite{Ziman2001}. This $T^2$ dependence would take the same form as Eq. \ref{n5}, with $n=2$, and where $\alpha_2$ is a constant coefficient describing the scattering rate of the $T^2$ behavior. We find the $T^2$ dependence at low temperature much more accurately reflects the temperature dependence of resistivity [inset in Fig. \ref{F2}(a)].

The magnetoresistance (MR), defined as [$\rho(B)$-$\rho(B=0)$]/$\rho(B=0)$, for $I$ along [001] and the magnetic field $B$ along [110] between 1.8 and 30 K, is shown in Fig. \ref{F2} (c). The MR in CoTe$_2$ is extremely small but at lower temperatures $T<10$ K it is linear in magnetic field, and non-saturating. The linear behavior gradually changes with temperature. A $B^{1.31}$ dependence is observed at 10 K, $B^{1.44}$ dependence at 20 K, and finally exhibiting a more conventional $B^2$ dependence at 30 K. We verified this behavior in more than one samples from different growth batches (see Fig. S6).The linear MR is an interesting property and has been observed in many material systems \cite{Zhang2011,Wang2012,Khouri2016,Kozlova2012,Xu1997,Zhao2015,Feng2015,Novak2015,Tang2011,Gusev2013,Wang2012}. However, it is frequently observed and used as evidence for materials exhibiting topological features in their electronic structure \cite{Abrikosov1998,Hu2008,Zhang2011,Wang2012,Zhao2015,Fang2015,Novak2015,Tang2011,Gusev2013,Leaahy2018}.  One prevalent non-topological reason for the linear MR is disorder \cite{Xu1997,Kozlova2012,Hu2008,Parish2003}. But in such a case, the linear MR  usually spans a large temperature range unlike the quickly attained quadratic behavior at 30 K in CoTe$_2$. Another widely observed reason for the linear MR is an admixture of a component of the Hall resistivity (which depends linearly on the magnetic field) to the longitudinal resistance caused by inhomogeneities such as density or sample thickness variations \cite{Bruls1981}. Such variations give rise to a gradient of the (transverse) Hall voltage in the longitudinal direction that will naturally be picked up in measurements of the longitudinal resistance, therefore giving rise to a linear component in the magnetoresistance. In CoTe$_2$, the Hall resistivity is linear at all temperatures measured [Fig. \ref{F2}(d)] and its influence only at the lower temperatures, where the MR is actually larger, is not consistent with this mechanism. In fact, thermal conductivity measured as a function of temperature (see Fig. S7) shows a sharp drop below about 30 K indicating that Umklapp process is frozen out below this temperature and thus the effect of phonon scattering is smaller at the temperature range where linear MR is observed, consistent with the conclusion drawn from the analysis of the heat capacity data. This observation is indicative that with the increase in temperature phonons may play a role in suppressing the linear MR in  CoTe$_2$. As the  magnetotransport of such a system with the coexisting nodal points, lines and planes have never been measured and analyzed before, this anomalous magnetotransport behavior of CoTe$_2$ calls for further theoretical and experimental investigations. 

\begin{figure}[!ht]
\begin{center}\includegraphics[width=1\linewidth]{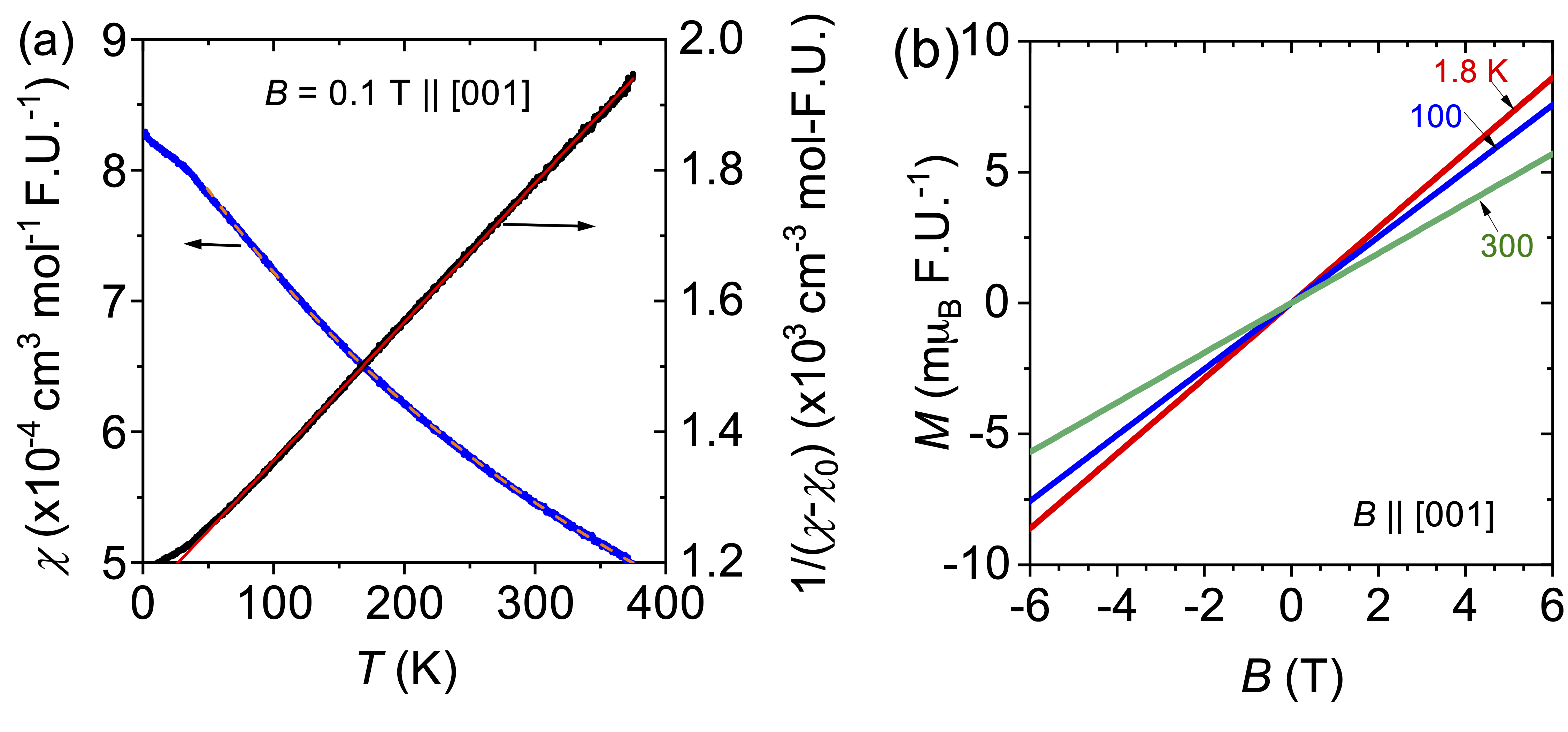}
\end{center}
\caption{ (a) DC magnetic susceptibility ($\chi=M/B$) as a function of temperature for $B = 0.1$ T along the $c$-axis (left axis). The dashed line shows a fit to the Curie-Weiss law between 50 and 375 K, yielding $\chi_0=-1.585\times10^{-5}$ cm$^3$ mol$^{-1}$ F.U.$^{-1}$. 1/($\chi-\chi_0$) vs $T$ is plotted in the right axis (black line). The red solid line is a Curie-Weiss fit. (b) Magnetization vs magnetic field at the indicated temperatures.}
\label{F3}
\end{figure}

\begin{figure}[!ht]
\begin{center}\includegraphics[width=1\linewidth]{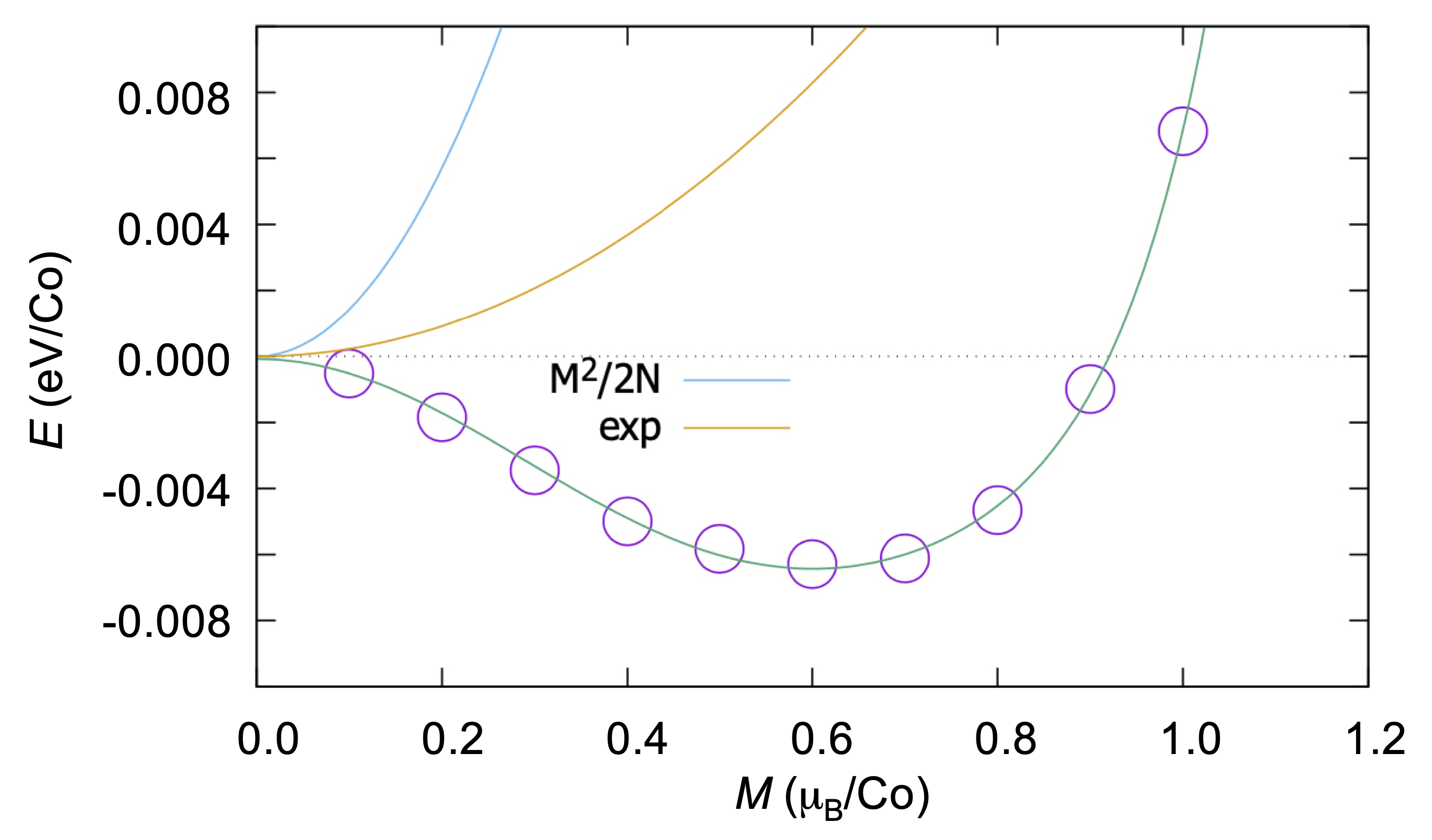}
\end{center}\caption{
DFT calculation predicting a ferromagnetic ground state (purple points). The blue line is the parabola corresponding to the Pauli susceptibility with the calculated density of states $N$. The orange line is obtained from the Moriya's SCRT formula with the spin fluctuations amplitude adjusted to match the experimental zero temperature susceptibility (see the text for more details).}
\label{F4}
\end{figure}

To characterize the magnetism in CoTe$_2$ we performed DC magnetic measurements. The blue curve in Fig. \ref{F3}(a) (plotted in the left axis)  shows magnetic susceptibility as a function of temperature between 1.8 and 375 K. The susceptibility increases slowly with decreasing temperature but does not show a sign of a long-range ordering, consistent with resistivity and heat capacity measurements. It neither diverges at lower temperatures as in normal paramagnetic materials. It is an indication that the material tends to order but the ordering is somehow suppressed (by strong spin fluctuations, which we will discuss below). Inverse susceptibility after subtracting a small diamagnetic background ($\chi_0$ of $-1.58\times$10$^{-5}$ cm$^{3}$mol$^{-1}$F.U.$^{-1}$) determined from the Curie-Weiss fit to the susceptibility above 50 K (indicated by the red dashed line) is plotted in the right hand axis of Fig. \ref{F3}(a). Thus obtained  inverse susceptibility is linear over a wide range of temperature (down to 25 K) and its fitting to the Curie-Weiss law [red solid line in Fig. \ref{F3}(a)]:
\begin{equation}
{1}/({\chi-\chi_0})=({T}-{T_\theta})/{C},
\end{equation}
where $\chi$ is the spin susceptibility, $C$ is the Curie constant, and $T_\theta$ is the Curie-Weiss
temperature, yields the effective moment ($\mu_{\mathrm{eff}}$=$\sqrt{8C}$) of 1.940 $\mu_{\mathrm{B}}$ per Co. This agrees reasonably well with the theoretical $\mu_{\mathrm{eff}}$ of 1.732 $\mu_{\mathrm{B}}$ per Co for the low-spin state ($S=1/2$) Co$^{4+}$. 
$T_\theta$ obtained from the $x-$intercept is $\sim 30$ K. The sizable positive value indicates the presence of ferromagnetic correlations. Magnetization data ($M$ vs $B$) depicted in Fig. \ref{F3}(b) is linear at all temperatures between 1.8 and 300 K indicating that the paramagnetism observed at room temperature, unusually, persists down to the lowest measured temperatures. The fact that despite the large $T_\theta$ the material does not order down to $\sim 1 $ K indicates strong spin fluctuations.

Additional insight can be gained from the density functional (DFT) calculations.  
Being mean-field by nature, DFT calculations do not account for fluctuation-driven suppression of magnetic order. Correspondingly, calculating the ferromagnetic total energy in DFT as a function of spin magnetization
yields a ferromagnetic ground state with a moment of $M= 0.6\mu_{\mathrm{B}}$/Co, 
also corresponding to $S=1/2$ (with a typical for DFT hybridization reduction), whose total stabilization energy is however only $\sim$6 meV/Co), as shown in Fig. \ref{F4}. The effect of  fluctuations can be {approximated } by means of Moriya's SCRT\cite{Moriya1985, Larson2004}.  If the DFT
energy is {expanded as a } polynomial in $M$,
\begin{equation}
E=a+bM^2+cM^4+dM^6+eM^8...,
\end{equation}
and there are Gaussian fluctuations of the magnetic moment with an average
squared amplitude $\xi^2$, then, in SCRT, the renormalized energy is given by
\begin{equation}
E=a+(b+5c\xi^2+35d\xi^4+35e\xi^6)M^2 + ...
\end{equation}
and the renormalized susceptibility is, as usually, $
    \chi^{-1}={\partial^2E}/{\partial M^2}$.
Typically, Moriya's parameter $\xi$ is comparable with, or somewhat larger 
than the equilibrium DFT value.
Fitting our DFT calculations (the green line in Fig. \ref{F4}), and assuming fluctuations with $\xi\approx 0.71$ $\mu_{\mathrm{B}}$, consistent with the DFT moment of $0.6\mu_{\mathrm{B}}$, we can reproduce the experimental susceptibility at $T = 0$. Thus, CoTe$_2$ is a material where ferromagnetic order is suppressed by strong spin fluctuations, leaving the ground state on the verge of a magnetic quantum critical phase transition. 

\section*{Conclusion}
To conclude, our measurements and calculations suggest that CoTe$_2$ sports a unique combination of a quantum criticality and multiple nodal Dirac lines. The ferromagnetic order is wholly suppressed, maintaining paramagnetic behavior down to the lowest measured temperatures due to  strong spin fluctuation. The calculated bands show that this material has potential for harboring new topological physics, as CoTe$_2$ hosts both nodal Dirac lines, and more conventional Dirac points directly on the Fermi surface, within  a single material on the verge of quantum criticality. This study should prompt future investigations into system, for instance, by probing the Dirac dispersions via photoemission and quantum oscillations.
A quantum critical point may be reached by doping or pressure to provide further insight into the interplay between magnetism and topological fermions.

\section*{Methods}

\textbf{Single crystal growth and x-ray diffraction.} Single crystals of CoTe$_2$ were grown by molten metal flux method with Te as a self flux.  Co powder (Alfa Aesar; 99.998\%), and Te shots (Alfa Aesar; 99.9999\%), were loaded in a 2-ml aluminum oxide crucible in a molar ratio of 1:19. The crucible was then sealed in a fused silica ampule under vacuum. The sealed ampule was heated to 800°C over 12 hours, homogenized at 800°C for 10 hours. The furnace was then quickly cooled to 760°C  in 2 hours followed by slow cooling to 500°C over 140 hours. Once the furnace reached 500°C, the excess flux was decanted from the crystals using a centrifuge. Well-faceted rectangular crystals as large as 30 mg were obtained. The crystal structure of the compound was first verified by x-ray powder diffraction at room temperature using a Rigaku MiniFlex diffractometer and then by single crystal x-ray diffraction. A few crystals from each growth batch were ground into powder, and x-ray diffraction patterns were collected on those powder samples. Rietveld refinement \cite{Mccusker1999} of  powder x-ray pattern was performed using FullProf software \cite{Rodriguez-carvajal1993}. Single-crystal X-ray diffraction data of CoTe$_2$ were collected on a piece of a representative single crystal at room temperature using a Rigaku XtaLAB Synergy-i diffractometer with the HyPix-Bantam direct photon-counting detector. The single-crystal was glued on a loop and mounted on the goniometer head of the diffractometer. The crystal structure of CoTe$_2$  was solved with space group $Pnnm$ and refined with the SHELX program \cite{Sheldrick}. 

\textbf{STEM Measurements.} An FEI Helios NanoLab 660 DualBeam (SEM/FIB) system was used to prepare cross-sectional TEM samples. Electron-beam-induced deposition of Pt was initially deposited on top of the materials to protect the sample surface, followed by 3-$\mu$m ion-beam induced Pt deposition. To reduce Ga ion damage, 2 kV Ga ion beam was applied to thin the lamella in the final step. An FEI Titan 80-300 STEM/TEM equipped with a probe spherical-aberration corrector was employed to conduct atomic resolved scanning transmission electron microscopy (STEM) imaging.

\textbf{Raman measurements.} Raman measurements of CoTe$_2$ single crystals were performed at room temperature in a backscattering geometry. The sample was excited by a 532 nm laser focused through a 0.75 NA objective lens with 40$\times$ magnification. The laser power was measured to be 200 $\mu$W before the objective. Scattered light was collected through the same lens and directed to an imaging spectrograph. Rejection of the laser is accomplished using a pair of long pass dielectric filters that allow measurement of Raman scattered light down to 75 cm$^{-1}$. 

\textbf{Magnetic, transport and thermal measurements.} Electronic transport measurements were conducted within the 9-T Quantum Design PPMS with the DC resistivity option. Samples were polished to dimensions of approximately 1.00$\times$0.40$\times$0.12 mm with the long axis corresponding to the [001] direction. An excitation current of 4 mA was used for all measurements. Electrical contacts were affixed using Epotek H20E silver epoxy and 25 $\mu$m Pt wires with typical contacts resistances of $\approx$ 10 $\Omega$, such that current was directed along the [001] direction. In magnetoresistance measurements, the contact misalignment was corrected by field symmetrizing (for MR) and antisymmetrizing (for Hall) the measured data. Magnetization measurements were made using a Quantum Design VSM SQUID in the field-cooled mode. Heat capacity was measured with the Quantum Design PPMS. The thermal conductivity was measured via the Quantum Design PPMS Thermal Transport Option (TTO) where the sample was polished to a rectangular bar and mounted with the standard TTO thermometer shoes. The sample geometry yielded a cross-sectional area of 0.747 mm$^2$ with thermometer lead separation of 1.628 mm. The applied heat current was directed along the c-axis of the sample with the heat pulses maintained at 3\% of the sample temperature for each temperature setpoint measured.

\textbf{First-principles Calculations.} Most calculations were performed using
the all-electron linearized augmented plane wave code WIEN2k\cite{WIEN2K} with gradient approximation for the exchange-correlation 
energy\cite{gga}, including the results shown in Figs. \ref{F1} and \ref{F3}.  For control purposes, some calculations were performed using projected augmented wave pseudo-potential code VASP\cite{vasp}. Fig. S4 was generated using the Full Potential  Local Orbitals (FPLO)\cite{fplo} package (agreement of the results from WIEN2k was verified).

\section*{Acknowledgements}
The authors thank  Binghai Yan for insightful discussions. N.J.G acknowledges the support from the NSF CAREER award DMR-2143903. Crystal growth part of the work at George Mason University was supported by the U.S. Department of Energy, Office of Science, Basic Energy Sciences, Materials Science and Engineering Division. I.I.M. acknowledges support from the U.S. Department of Energy through the grant No. DE-SC0021089. M.P.G. acknowledges the Alexander von Humboldt Foundation, Germany and IFW-Dresden, Germany for the equipment grants. R.G. thanks UGC-Nepal for the fellowship with award number MRS-77/78 S \& T -119. Magnetic characterization at G.U. have been supported by the NSF (ECCS-2151809). The acquisition of a Magnetic Property Measurements System (MPMS3, Quantum Design) at G.U. used herein was supported by the NSF (DMR-1828420). H.Z. acknowledges support from the U.S. Department of Commerce, NIST under financial assistance award 70NANB19H138. A.D. acknowledges support from the Material Genome Initiative funding allocated to NIST. J.J. and P.M.V. acknowledge support from the NSF CAREER award DMR-1847782.
\\
\section*{References}

%\bibliography{main.bib}

\begin{widetext}
\pagebreak
\renewcommand\thesection{\arabic{section}}
\renewcommand{\theequation}{S\arabic{equation}}
\setcounter{equation}{0}
\setcounter{figure}{0}
\setcounter{table}{0}
\setcounter{page}{1}
\makeatletter
\renewcommand\thesection{S\arabic{section}}
\renewcommand{\theequation}{S\arabic{equation}}
\renewcommand{\thetable}{S\arabic{table}}
\renewcommand\thefigure{S\arabic{figure}}
\renewcommand{\theHtable}{S\thetable}
\renewcommand{\theHfigure}{S\thefigure}

    \renewcommand{\arraystretch}{1.8}

\begin{center}
\textbf{\large Supplementary Information for: \\
\vspace{0.5cm}
CoTe$_2$: A quantum critical Dirac metal with strong spin fluctuations}
\\
\vspace{0.3cm}
Peter E. Siegfried, Hari Bhandari, Jeanie Qi, Rojila Ghimire, Jayadeep Joshi, Zachary T. Messegee, Willie Beeson, Kai Liu, Madhav Prasad Ghimire, Yanliu Dang, Huairuo Zhang, Albert Davydov, Xiaoyan Tan, Patrick M. Vora, Igor I. Mazin, and Nirmal J. Ghimire
\end{center}

\section{X-ray diffraction}
\begin{table}[H]
\caption{Structure Refinement Parameters, Atomic Coordinates  and Anisotropic Thermal Parameters for CoTe$_2$.}\label{T1}
\centering
\begin{tabular}{@{\hspace{.7cm}}l@{\hspace{.7cm}}l}	
 \hline
Formula&CoTe$_2$\\% [0.5ex] 
 Crystal system         &              		  Orthorhombic	$\hspace{5.0cm}$ \\ 
  Temperature (K)  & 293  \\ 
 $\lambda$ (\AA) & 0.71073 \\
  Space group & {\it Pnnm} \\
 Unit cell $a$, $b$, $c$ (\AA) & 5.3176(10),~6.3186(2),~3.8864(10)\\
 {\it V}, (\AA$^3$) & 130.582(6) \\ 
  {\it Z} & 2 \\ 
  Crystal size (mm$^3$) & 0.082, 0.126, 0.065 \\ 
 $\rho_{\mathrm{calc}}$, g cm$^{-3}$ & 7.989 \\ 
 $\mu$, mm$^{-1}$ & 28.074 \\ 
 $\theta_{\mathrm{max}}$, deg & 36.265 \\ 
 Reflections collected & 2277 \\ 
 $R_{int}$ & 0.0286 \\ 
 Unique reflections & 357 \\ 
 Parameters refined & 12 \\ 
 $R_1$,w$R_2$ [$F_0>4\sigma(F_0)$] & 0.0278, 0.0751 \\ 
 Diff. peak and hole, e (\AA$^{-3}$) & 2.227, –3.577 \\
 Goodness-of-fit & 1.192 \\%[1ex] 
\end{tabular}
\begin{tabular}{l@{\hspace{1.1cm}}l@{\hspace{1.1cm}}l@{\hspace{1.1cm}}l@{\hspace{1.1cm}}l@{\hspace{1.1cm}}l}								
		\hline	
  \hline
        Atom           & Wyck.Site      & $x$         &	  $y$	          &	      $z$	         &	  Occupancy	        \\
		\hline
 Co & 2$a$ & 0 & 0 & 0 & 1.00  \\														
 Te & 4$g$ & 0.21949(6) & 0.36340(5) & 0 & 1.00  \\ 
  \hline      					
	    \hline
\end{tabular}
\begin{tabular}{l@{\hspace{1.1cm}}l@{\hspace{1.1cm}}l@{\hspace{1.1cm}}l@{\hspace{1.1cm}}l@{\hspace{1.1cm}}l}																					
     U$_{11}$ & U$_{22}$ & U$_{33}$ & U$_{23}$ & U$_{13}$ & U$_{12}$\\
(\AA$^2$) & (\AA$^2$) & (\AA$^2$) & (\AA$^2$) & (\AA$^2$) & (\AA$^2$)\\
\hline
0.0094(4) & 0.0077(4) &	0.0091(4) &	0 &	0	& -0.0002(3) \\ 
0.0064(2) &	0.0057(2)	& 0.0075(2) &	0 &	0 &	-0.00073(7) \\    					
	    \hline
\end{tabular}

\label{T1}
%\end{center}
\end{table}

%********************************************************************************************

%*************************************************************************************************************************************************************

\begin{figure}[ht!]
\begin{center}
\includegraphics[width=6.5in]{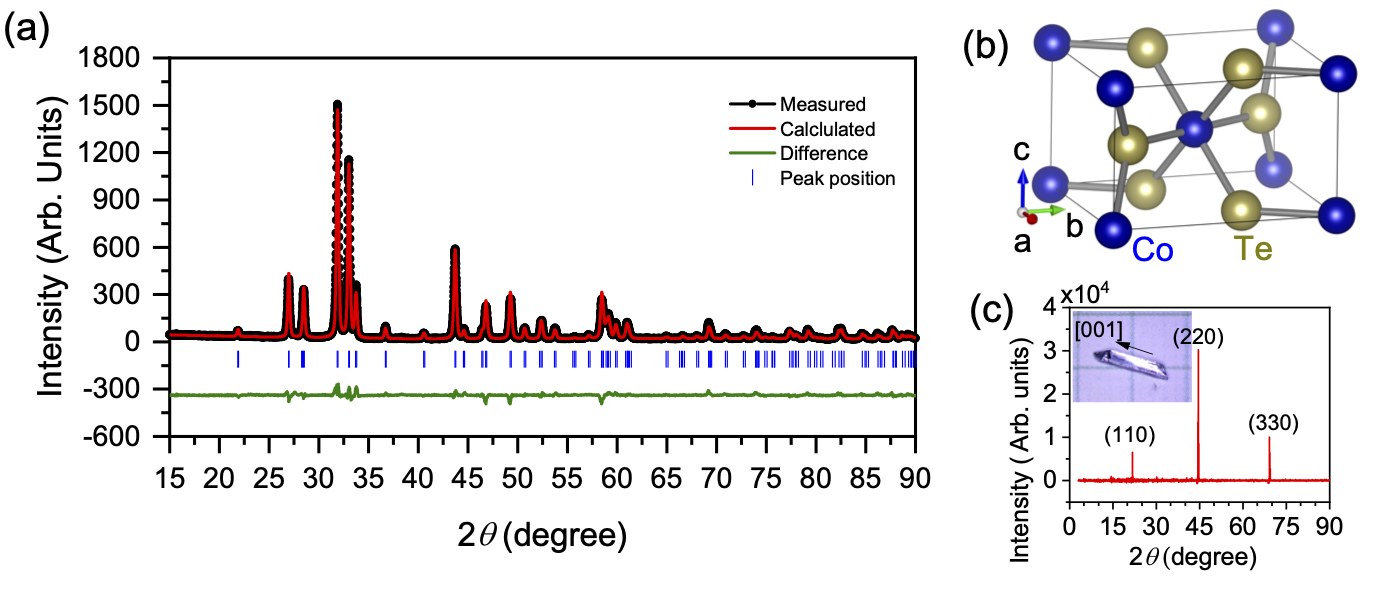}
\caption{ a) Rietveld refinement of x-ray powder pattern of CoTe$_2$ collected at room temperature. b) Crystal structure of CoTe$_2$. c) (hk0) x-ray peaks obtained from the as-grown flat face of a single crystal of CoTe$_2$ as shown in the inset. The [001] direction is along the length of the crystal as indicated by the arrow in the inset.}
\label{S1}
\end{center}
\end{figure}

\section{STEM Measurements}

\begin{figure}[ht!]
\begin{center}
\includegraphics[width=5.5in]{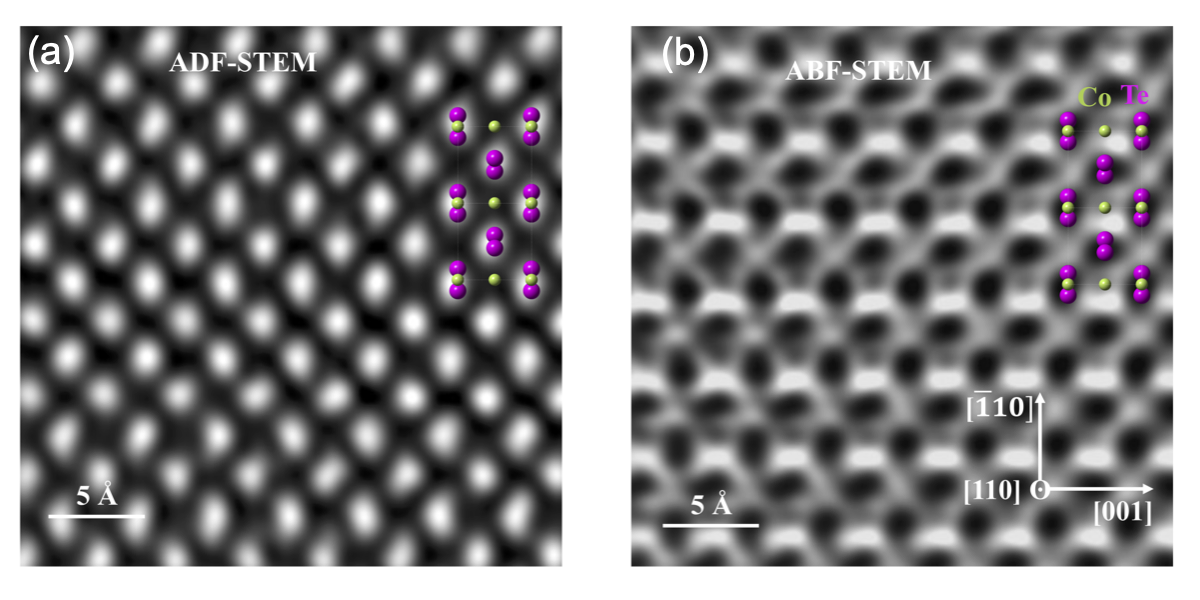}
\caption{ Scanning transmission electron microcopy (STEM) images of CoTe$_2$. (a) Annular dark-field STEM image. (b) Annular bright-field STEM image.}
\label{STEM}
\end{center}
\end{figure}

Figure \ref{STEM} shows the atomic resolved annular dark-field (ADF) and annular bright-field (ABF) STEM images which were taken simultaneously along the [110] zone-axis. The inserted atomic model of CoTe$_2$  with the $Pnnm$ space group matches well with the atomic images. 

\section{Raman Measurements}
\begin{figure}[ht]
\begin{center}
\includegraphics[width=3.5in]{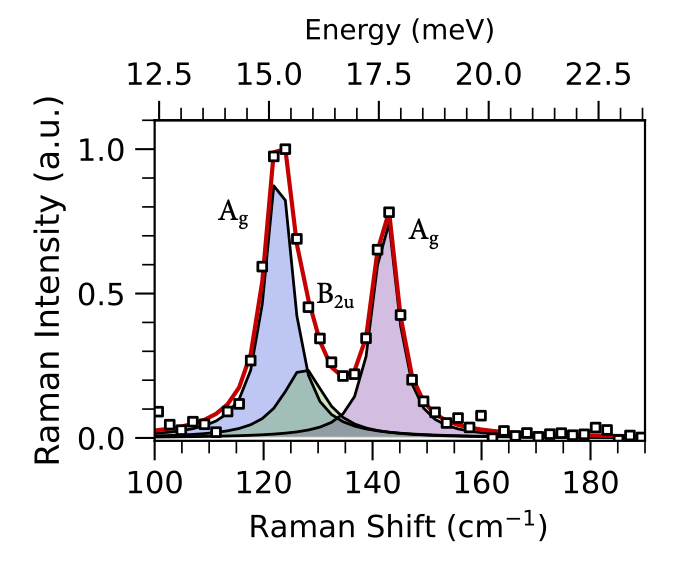}
\caption{ Room temperature CoTe$_2$ Raman spectrum (squares) fit with the sum of three Lorentzian functions (red line). Individual Lorentzians comprising the fit are shaded in blue, green, and purple.}
\label{RAMAN}
\end{center}
\end{figure}

Raman spectrum of CoTe$_2$ is presented in Fig. \ref{RAMAN}. The phonon mode frequencies are determined by fitting the data to a sum of three Lorentzians. The spectrum is dominated by two modes at 122.79 cm$^{-1}$ and 142.41 cm$^{-1}$ to which we assign Ag symmetry based on our DFT calculations (Table \ref{T3}). There is additional spectral weight between these two modes that can be accounted for with a third Lorentzian centered at 127.32 cm$^{-1}$. While no Raman active modes are expected here, DFT calculations do predict a mode of B$_{2u}$ symmetry. We therefore suggest this feature is a B$_{2u}$ symmetry mode activated by disorder. 

%********************************************************************************************
\begin{table}[h]
\caption{Experimental and Theoretical Phonon Frequencies Determined from Fits to the Raman Spectrum in Figure \ref{RAMAN} and DFT Calculations, Respectively.} \label{T3}
\centering
\begin{tabular}{|l|c|c|c|c|}
 \hline
 \multicolumn{2}{|c|}{Experiment} &
  \multicolumn{2}{c|}{DFT Calculation} \\
 \hline																					
Mode Frequency (cm$^{-1}$) & Symmetry & Mode Frequency (cm$^{-1}$) & Symmetry \\
\hline
	122.79 $\pm$ 0.15 & 	A$_g$	& 127.4524 $\pm$ 0.9480	 &  A$_g$	   \\
	127.32 $\pm$ 1.43 & 	B$_{2u}$	& 130.7357 $\pm$ 0.6270	 & B$_{2u}$  \\
	142.41 $\pm$ 0.08 & 	A$_g	$ 	&  144.7070 $\pm$ 0.0288 & 	A$_g	$\\
\hline
\end{tabular}

\end{table}
%*************************************************************************************************************************************************************
\section{First-principles Calculations}
 Figure \ref{SDFT}(a) shows the band structure calculated with (red lines) and without (blue lines) spin-orbit coupling (SOC) [the latter is the same as the one presented in Fig. 1(a) in the main text]. Bands along the maximum-splitting directions are shown in Figs. \ref{SDFT}(b-k) from the corresponding points labelled as b-k in Panel (a) showing how the band degeneracy is lifted when moved away from these points forming the nodal lines at the Fermi surface depicted by the red and blue lines in Fig. 1(b) in the main text. It is to be noted that the Dirac point between S and R labelled by i, j in Panel (a) is a special point where the bands remain degenerate along $k_x$, $k_y$, and $k_z$ as shown in Fig. \ref{SDFT}(i) (moving along $k_y$; same behavior is observed moving along $k_x$, and $k_z$ that are not shown here). The the degeneracy is lifted in-between as shown by a representative plot calculated from (0.3, 0.5, 0.14) to (0, 0, 0.25) and schematically illustrated by the four-fold symmetry in Fig. \ref{CoTe2Dirac}. The Dirac points b, c and i in Fig. \ref{SDFT}(a) [see Figs. \ref{SDFT}(b, c and i)] are symmetry protected and do not open gap with SOC, consistent with the prediction of ref. 13. SOC opens gaps at all other Dirac points indicated in Fig. \ref{SDFT}(a) [see Figs. \ref{SDFT}(d, e, f, g, h, j, and k)].

\begin{figure}[!ht]
\begin{center}
\includegraphics[width=.95\linewidth]{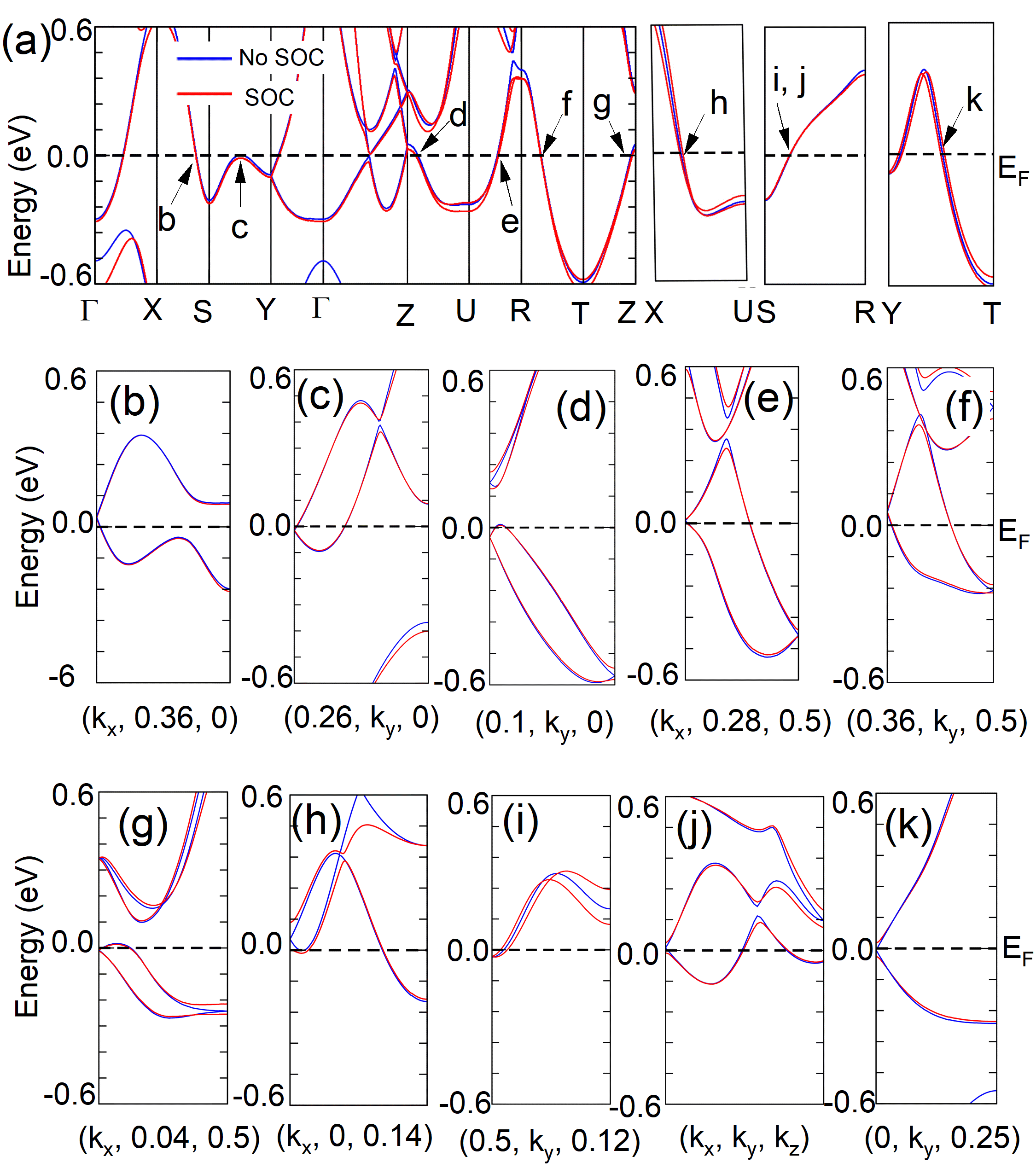}
\end{center}
\caption{ (a) Band Structures of CoTe$_2$ without SOC (blue) and with SOC (red). (b-k) Bands along $k_x$, $k_y$, and $k_z$ from the Dirac points indicated by letters b to k in panel (a) showing the lifting of degeneracy along the directions as indicated. Point i in panel (a) is a special point where the degeneracy is protected along $k_x$, $k_y$, and $k_z$ [variation along $k_y$ is shown in panel (i)], and lifted in between as show shown in panel (j) where all $k_x$, $k_y$, and $k_z$ are varied simultaneously.}
\label{SDFT}
\vspace{1cm}
\end{figure}

\begin{figure}[!ht]
\begin{center}
\includegraphics[width=.3\linewidth]{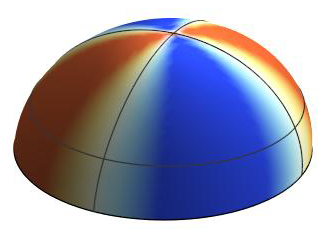}
\end{center}
\caption{Schematic of the degeneracy of Dirac points along SR [see Fig. \ref{SDFT}(a)]. The degeneracy is protected along $k_x$, $k_y$, and $k_z$, but lifted at other points showing a four-fold degeneracy-lifting pattern. The diagram shows a patch of the Fermi surface around the point (top of the figure) where it intersect with the SR line (vertical in the cartoon). The color intensity represents the amplitude of the Dirac splitting, and different colors indicate that the band character is swapped by crossing a nodal line. }
\label{CoTe2Dirac}
\end{figure}

\section{Magnetoresistance and Thermal Conductivity}

\begin{figure}[ht!]
\begin{center}
\includegraphics[width=.54\linewidth]{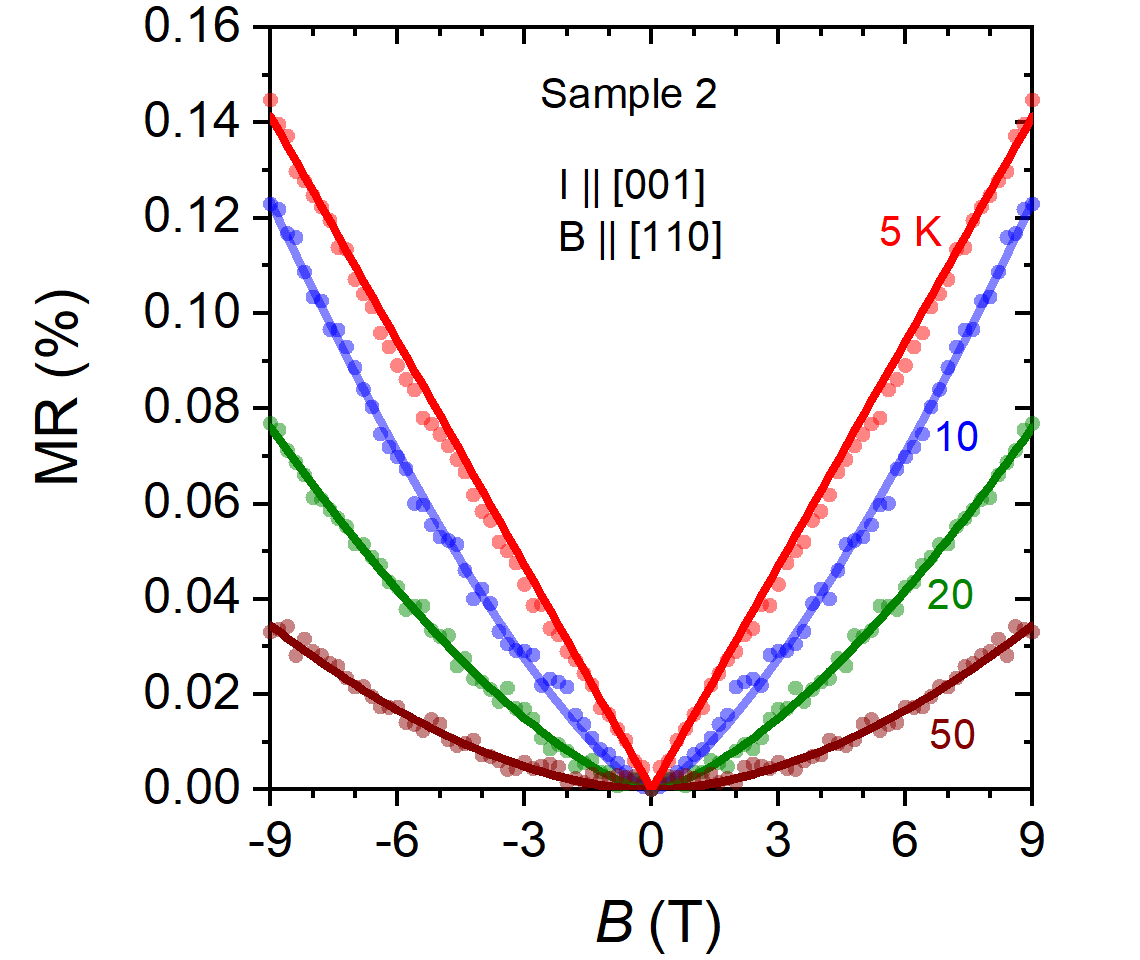}
\end{center}
\caption{ Magnetoresistance of a second sample from a growth batch different from that shown in Fig. 2(c) of the main text showing the linear MR at 5 K, which gradually deviates from the linear behavior. The solid spheres are the measured data, and the solid lines are $B^{n}$ fits to the measured data, where $B$ is the magnetic field. At 5 K, $n =1$  which gradually increases toward 2 with the increase in temperature.}
\label{MR2}
\end{figure}

\pagebreak
\begin{figure}[ht!]
\begin{center}
\includegraphics[width=.54\linewidth]{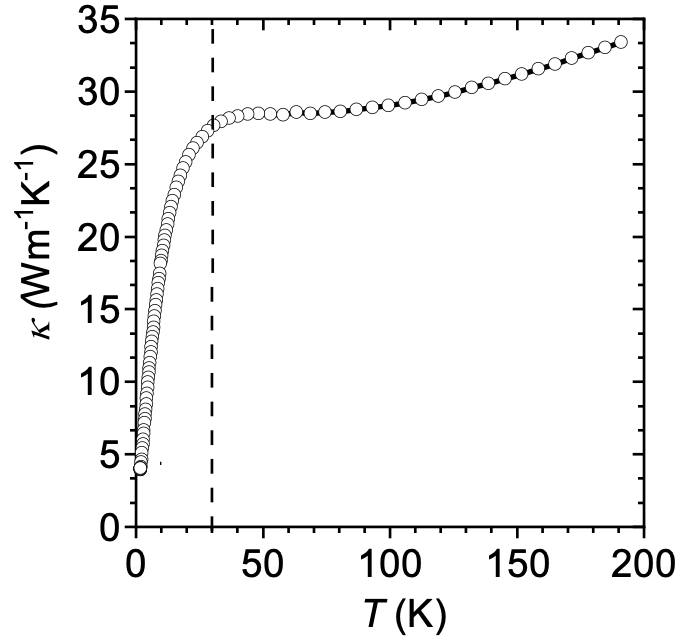}
\end{center}
\caption{ Thermal conductivity ($\kappa$) of CoTe$_2$ as a function of temperature. The dashed vertical line is a guide to eye for T = 30 K.}
\label{TC}
\end{figure}

\end{widetext}

\end{document}